\documentclass{aa}
\usepackage{graphics}
\usepackage{psfig}
\begin{document}
\title{Atomic and molecular gas in the merger galaxy NGC~1316 (Fornax~A) 
and its environment}
\subtitle{}
\author{C. Horellou
\inst{1}
\and
J.H. Black 
\inst{1}
\and
J.H. van Gorkom
\inst{2}
\and 
F. Combes
\inst{3}
\and
J.M. van der Hulst
\inst{4}
\and 
V. Charmandaris
\inst{5}
          }

\offprints{horellou@oso.chalmers.se}
\institute{Onsala Space Observatory, Centre for Astrophysics \&\ Space Science,
Chalmers University of Technology, 
{S--439~92} Onsala, Sweden  \\
\email{horellou@oso.chalmers.se, jblack@oso.chalmers.se}
\and
Astronomy Department, Columbia University, 538 W 120 Street, New York, NY 10027, 
USA \\  
\email{jvangork@astro.columbia.edu}
\and
Observatoire de Paris, DEMIRM, 61 avenue de l'Observatoire, F-75014 Paris, France \\
\email{Francoise.Combes@obspm.fr}
\and
Kapteyn Institute, Department of Astronomy, 
Postbus 800, 9700 AV Groningen, The Netherlands \\
\email{j.m.van.der.hulst@astro.rug.nl}
\and
Cornell University, Astronomy Department, 
106 Space Sciences Building, 
Ithaca, NY 14853, USA \\
\email{vassilis@astro.cornell.edu}
             }
\date{Received 4 May 2001/ Accepted 18 July 2001}
\abstract{ 
 We present and interpret 
 observations of atomic and molecular gas toward the southern elliptical galaxy
 NGC~1316 (Fornax~A), a strong double-lobe radio source with a disturbed optical morphology 
 that includes numerous shells and loops. 
 The $^{12}$CO(1--0), $^{12}$CO(2--1), and H{\sc i} observations 
 were made with SEST and the VLA. 
 CO emission corresponding to a total molecular hydrogen mass of $\sim$5$\times$10$^8$ M$_\odot$ 
 was detected toward the central position as well as northwest and  
 southeast of the nucleus in the regions of the dust patches.   
 The origin of that gas is likely external 
 and due to accretion of one or several small gas-rich galaxies. 
 H{\sc i} was not detected in the central region of NGC~1316, but 
 $\sim$2$\times$10$^7$ M$_\odot$ 
 of atomic gas was found 
 toward the giant H{\sc ii} region 
 discovered by Schweizer (1980)
 located $6{\farcm}7$ (or 36.2 kpc) from the nucleus.  
 H{\sc i} was also found at three other locations in the outer part of NGC~1316. 
 The H{\sc i} distributions and kinematics of the two nearby spiral companions of NGC~1316, 
 NGC~1317 (a barred galaxy to the north) and 
 NGC~1310 (to the west) could be studied. Both galaxies have unusually 
 small H{\sc i} disks that may 
 have been affected by ram-pressure stripping. 
   }
\titlerunning{Atomic and molecular gas in and around NGC~1316}
\maketitle
\def\hi{H{\sc i\,}}
\def\hii{H{\sc ii\,}}
\def\kms{km~s$^{-1}$}

\section{Introduction}

\begin{figure*}
\psfig{file=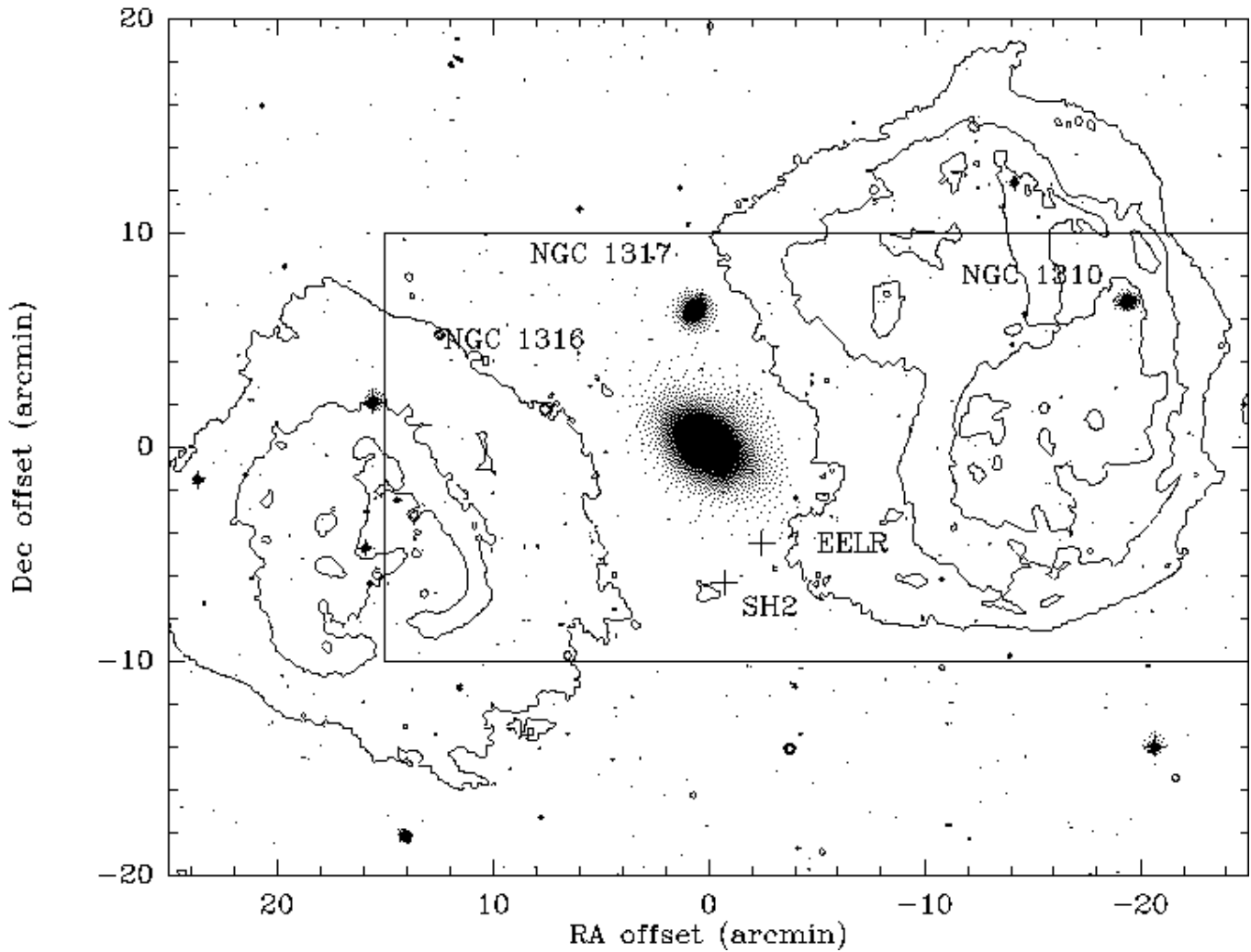,width=18cm,angle=-90}
\caption{ 
Radio continuum contour map at  $\lambda$20 cm of Fomalont et al. (\cite{fomalont89})  
superimposed on an optical image taken from the Digitized 
Sky Survey. The resolution of the radio continuum map is 14$''$. 
The two neighbor galaxies, NGC~1317 and  NGC~1310, and the location of 
the extended emission line region EELR and the giant \hii region SH2  
in the southern tidal loop called L$_1$ by Schweizer (\cite{schweizer80}) 
are indicated. NGC~1310 is located in the foreground
of the western radio lobe.  
The black rectangle shows the size of the image displayed in Fig.~2. 
}
\end{figure*}

Over the past fifteen years, evidence 
for the existence of a complex, multi-phase 
interstellar medium in elliptical galaxies 
has accumulated. 
Most of the gas in ellipticals is hot 
(10$^7$ K) and radiates at X-ray wavelengths, but there is also a cooler
component of \hi, CO, ionized gas and dust 
(e.g. Bregman et al. \cite{bregman92}). 
The origin and fate of the gas
are still debated, but it seems that the hot gas is associated with
the spheroid, whereas the atomic and molecular gas as traced by the CO 
characterize a prominent disk
(e.g. van Gorkom \& Schiminovich {\cite{vangorkom97}). 
The detection rates of \hi or CO are significantly
higher among ellipticals with peculiar features 
(such as tidal tails, loops, shells, dust lanes) than in the 
whole class of ellipticals, which suggests that the gas is
of external origin and due to accretion of smaller companions and/or 
to major mergers. 
 
NGC~1316 (Fornax~A, Arp~154, PKS~0320--37) 
on the outskirts of the Fornax cluster presents several characteristics
of a merger remnant. 
The disturbed outer morphology with numerous shells and loops was first
noted by Murdin et al. (\cite{murdin79}). The galaxy 
was extensively studied in the optical by Schweizer 
(\cite{schweizer80}, 
hereafter S80) and Schweizer (\cite{schweizer81}) who 
classified it as a D-type galaxy with an elliptical-like
spheroid surrounded by a large envelope.  
The brightness profile approximately obeys the $r^{1/4}$ law. 
The galaxy has an inner dust lane of length $\sim$2$'$ 
oriented close to the minor axis. 
S80 also reported the discovery of a giant H{\sc ii} region 
(hereafter denoted SH2) 
$6{\farcm}7$ from the nucleus, and the existence of an inner disk 
of ionized gas probably associated with the dust lane and 
extending out to 54$''$. 
The disk is in rapid rotation 
($v\, {\sin}i \sim 350$ \kms).
To explain those features and the lack of a nearby perturber, 
S80 suggested that NGC~1316 accreted one or several gas-rich
companion galaxies about $4 \times$ 10$^8$-- 2 $\times$ 10$^9$
years ago.  
Mackie \& Fabbiano (\cite{mackie98}) reported the discovery of an extended 
emission-line region (hereafter denoted EELR)
located in a tidal tail $5{\farcm}3$ from the nucleus
and suggested on the basis of its position and extreme size that it is related
to the merger process. 

NGC~1316 is the 
third nearest strong radio galaxy in the sky after NGC~5128 (Centaurus~A) and 
M~87 (Virgo~A) (see Fig.~1). 
The radio emission arises in two giant lobes  
separated by 33$'$  
and located mainly outside the optical galaxy, at a position angle of 
$\sim 110\degr$ 
(Wade \cite{wade61}; Ekers et al. \cite{ekers83}, 
Fomalont et al. \cite{fomalont89}).
In the central arcminute a bent radio jet 
abruptly terminates at the location of a dust lane, which may have deflected
it
(Geldzahler \& Fomalont \cite{geldzahler84}).
The nucleus is a bright UV source, unresolved by {\it HST}
(Fabbiano et al. \cite{fabbiano94}). 

A study of the stellar kinematics of NGC~1316 showed that the galaxy is rotating 
rapidly around the minor axis at a position angle of $\sim 140\degr$ 
(Bosma et al. \cite{bosma85}).
More recently Arnaboldi et al. (\cite{arnaboldi98})
measured radial velocities of 43 planetary nebulae in the outer parts 
of NGC~1316
to determine the distribution of angular momentum.
They derived a constant velocity $v$= 110 \kms\ along
the major axis of the galaxy in the radial range $R$=50--200$''$. 
They also probed the kinematics of the  inner region ($R <80''$) 
using absorption line spectra.
They confirmed the alignment between the photometric and the kinematic minor axes
but also found evidence for a disturbed kinematical pattern 
beyond $R \sim 30''$. 

X-ray emission from NGC~1316 has been observed with several instruments.  
The $L_X/L_B$ ratio of NGC~1316  is low compared to that of E and S0 galaxies
($Einstein$ observations by Fabbiano et al. \cite{fabbiano92}; 
Kim et al. \cite{kim92}),
but in the same range as that of dynamically young ellipticals which may have 
experienced mergers (Fabbiano \& Schweizer \cite{fabbiano95};
see also Sansom et al. \cite{sansom00}). 

In order to place NGC~1316 into the merger sequence as discussed by 
Hibbard \& van Gorkom (\cite{hibbard96}) it is necessary to know
the distribution of the cool gas, both the neutral atomic hydrogen
and the molecular hydrogen as traced by the CO. 
Sage \& Galleta (\cite{sage93}) detected $^{12}$CO(1--0) line emission 
at two positions along the inner dust lane and with kinematics similar to that of the
ionized gas. 
In this paper, we present 
more extensive and more sensitive maps of the $^{12}$CO(1--0) and the first 
$^{12}$CO(2--1) observations 
in the central part of NGC~1316 
as well as a search for CO emission toward the outer H{\sc ii} complexes. 
We also present results from a VLA search for \hi gas in NGC~1316 and its surroundings. 

The paper is organized as follows: 
We first describe the \hi and CO observations, then 
discuss the \hi results for the companion galaxies 
NGC~1317 and NGC~1310. 
We then focus on the \hi and CO results for NGC~1316 and the implications
for the merger picture. 

\section{Observations and data reduction}

\subsection{\ion{H}{i} observations}

The \hi observations consist of data collected with the Very Large Array
(VLA)\footnote{The VLA is a facility of the U.S. National Radio Astronomy Observatory,
which is operated by Associated Universities Inc. under contract with the U.S.
National Science Foundation}
in its spectral-line mode. 
The galaxy was observed in the hybrid DnC configuration: 
the 1.3 km D-array configuration with the antennas of the north arm in the
3 km C-array configuration to improve the North-South angular resolution, thus
compensating for the low declination of NGC~1316. 
The observations consisted of two runs of 5 hours each.  
 
A 6.25 MHz bandwidth was used, corresponding to a total velocity range 
of 1300 \kms. 
On-line Hanning smoothing has been applied to the data after which every other channel 
was discarded, leaving a set of 31 independent channels centered at 1774 \kms\  
and spaced by 41.7 \kms.
The observing parameters are summarized in Table~1. 

The data were inspected, calibrated, and maps were produced using NRAO's Astronomical
Image Processing System (AIPS). 
The continuum was estimated from channels free of line emission and was subtracted
in the UV plane. 
Line channel maps were made using uniform weighting with a ROBUST factor of 1 which, 
after some experimentation, was found to give the best combination of 
sensitivity and spatial resolution. 
The channel maps were CLEANed and a correction for primary beam attenuation was applied
to the entire data cube. 
Maps of the atomic hydrogen column density and velocity field were produced. 
The maps were not smoothed. Only pixels above a flux cut-off of 1.5 mJy/beam 
(2.5 $\sigma$) were included in the summation of the channel maps.  

The data were taken in 1985 by van Gorkom, Ekers, Schweizer and Wrobel. 
Due to limitations in on-line and off-line
computing power only 25 antennas and one polarization (RR) could be used.
The data were fully reduced by van Gorkom, but never published. Although NGC~1310,
NGC~1317 and the H{\sc ii} region were unambiguously detected, the quality
of the data reduction was  poor. This paper shows that reprocessing
of ancient (but venerable) data is very worthwhile, especially when a large
spectral dynamic range is needed, because there have been major improvements 
in software such as the UVLIN program within AIPS and far better weighting 
schemes in the imaging tasks.
\begin{table}
 \def\hf{\hfill}
 \caption[]{VLA observing parameters}
 \label{table1}
 \halign
{#&# &#\cr 
 \hline
 \noalign{\smallskip}
\hf	&\hf	 \cr
Phase center (B1950)\hf 		& $03^{\rm h}20^{\rm m}47{\fs}2$; 
					$-37^{\circ}23'08{\farcs}2$\hf \cr
Velocity center\hf			& 1774 km s$^{-1}$ (heliocentric)\cr
Primary beam (FWHM)\hf			& 30$\arcmin$\cr
Flux calibrator	\hf			& 3C 48\hf\cr
Phase calibrator\hf			& 0332--403\cr
Bandpass calibrator\hf			& 3C 48\hf\cr
Array \hf				& DnC\cr
Date \hf				& 19, 21 Oct 1985\hf\cr
Synthesized beam\hf\cr
-- FWHM: major $\times$ minor axis \hf 	& 52$'' \times 41''$ \hf\cr
-- Position angle\hf 			& +2$^{\circ}$\hf\cr
Bandwidth	\hf			& 6.25 MHz\hf\cr
Number of channels\hf			& 31\cr
Channel separation\hf			& 41.7 km s$^{-1}$\hf\cr
Time on source \hf			& 6 hours 51 min\hf\cr
Noise level (1$\sigma$)$^*$\hf		& \cr
- Flux density\hf			&0.6 mJy beam$^{-1}$ch$^{-1}$\hf\cr
- Column density\hf 			& 1.3$\times$10$^{19}$ cm$^{-2}$ beam$^{-1}$ch$^{-1}$\hf\cr
- \hi mass\hf 				& 2$\times$10$^6$ M$_\odot$ beam$^{-1}$ch$^{-1}$\hf\cr
 \noalign{\smallskip}
 \hline
 \noalign{\smallskip}
}
* Noise level in the continuum-subtracted maps before primary beam correction 
\end{table}

\begin{table*}
 \def\hf{\hfill}
 \caption[]{Basic parameters of the galaxies and outer \hii regions}
 \label{table2}
 \halign
{#	&# 	&# 		&# 		&# 		&#  &#\cr 
\hf 	&Note 	&NGC 1316\hf 	&NGC 1317\hf 	&NGC 1310\hf 	&SH2&EELR\cr
 \noalign{\smallskip}
 \hline
 \noalign{\smallskip}
RA(1950)\hf	&(1)	&03$^{\rm h}$20$^{\rm m}$47.2$^{\rm s}$ 
			&03$^{\rm h}$20$^{\rm m}$49.8$^{\rm s}$ 
			&03$^{\rm h}$19$^{\rm m}$08.8$^{\rm s}$
			&03$^{\rm h}$20$^{\rm m}$43.5$^{\rm s}$
			&03$^{\rm h}$20$^{\rm m}$34.8$^{\rm s}$\cr 
DEC(1950) 	&(1)	&--37$^\circ$23$'$08$''$					
			&--37$^\circ$16$'$50.9$''$
			&--37$^\circ$16$'$42.3$''$
			&--37$^\circ$29$'$28$''$
			&--37$^\circ$27$'$39$''$\cr
$v_{\rm helio}$ [km s$^{-1}$] &(2)	
			&1783	&1948	&1739	&1690	&\cr  
$D$ [Mpc]\hf	&(3)	&18.6	&18.6	&18.6	&18.6	&18.6\cr
Type 		&(4)	&.LXSOP.&.SXR1..&.SAS5*.\cr
Size		&(5)	
			&12$'\times8\farcm$5
			&2$\farcm75\times2\farcm$4 
			&2$\farcm0\times1\farcm$55
			&1$\farcs5\times3\farcs$5
			&81$''\times27''$\cr
P.A.  [$^\circ$]\hf &(5)	&50	&78	&95	&	&$\sim$45\cr 
$B_T^0$ [mag]\hf	&(6)	&9.4	&11.81	&12.56	&\cr
$L_B$ [L$_\odot$]\hf 	&(6)	&9.35$\times$10$^{10}$
			&1.02$\times10^{10}$
			&5.1$\times$10$^9$  
			&\cr
$L_{FIR}$ [L$_\odot$]\hf&(7)	&2.05$\times$10$^9$
				&2.45$\times$10$^9$
				&7.66$\times$10$^8$\cr
$M$(\hi) [M$_\odot$]\hf	&(8)	&$<$10$^8$ 
				&2.7$\times$10$^8$
				&4.8$\times$10$^8$
				&1.9$\times$10$^7$
				&$<$3$\times$10$^6$ \cr
$M$(H$_2$)  [M$_\odot$]\hf&(9)	&$\sim$5$\times$10$^8$ 
				&3.1$\times$10$^8$
				&$<$0.2$\times$10$^8$
				&$<$7$\times$10$^6$
				&$<$7$\times$10$^6$\cr
$v_{\ion{H}{i}}, \Delta v_{\ion{H}{i}}$ [km s$^{-1}$]\hf 	&(8) 
				&	
				&1960, 132 
                                &1827, 144 
                                &1690, $\leq 42$
				&1750, $\leq 42$ \cr
$F_{\ion{H}{i}}$ [Jy km s$^{-1}$]\hf 	&(8) 	
				&$<$0.1\hf
				&3.3$\pm$0.5
                                &$5.9\pm 1.3$
                                &0.2
				&0.1\cr
$v_{{\rm CO}}, \Delta v_{{\rm CO}}$ [km s$^{-1}$]\hf	&(9) 
				&	
				&1948, 84\hf 
				&\cr

\hf	&\hf	 \cr
 \noalign{\smallskip}
 \hline
 \noalign{\smallskip}
}
Notes: 	(1) NGC 1316: Schweizer (\cite{schweizer81});
		NGC 1317: da Costa et al. (\cite{dacosta98});
		NGC 1310: Loveday (\cite{loveday96});
		SH2 and EELR: calculated from the offsets relative to the central
		position given by Mackie \& Fabbiano (\cite{mackie98}). 
 
	(2) NGC 1316: Arnaboldi et al. (\cite{arnaboldi98}); 
		NGC 1317: Horellou et al. (\cite{horellou95});
		NGC 1310: RC3;
		SH2: this work.

	(3) Distance estimate for NGC 1365 based on {\it HST} measurements of 
		cepheid variable stars
		(Madore et al. \cite{madore99}). 
		At that distance, 1$'$ corresponds to 5.4 kpc. 
		That distance was adopted for all the systems discussed here.  

	(4) RC3.

	(5) For the three galaxies: optical diameters at the 25 mag arcsec$^{-2}$ isophote taken from RC3; 
		SH2: Schweizer (\cite{schweizer80});
		EELR: Mackie \& Fabbiano (\cite{mackie98}).
 
	(6) RC3.

	(7) Derived from the fluxes at 60 and 100 $\mu$m listed in the IRAS
	catalogue using the relation: 
$L_{FIR}= 3.94\times10^5 D^2 [2.58 f_{60} + f_{100}]$ 
where $D$ is the distance in Mpc, $f_{60}$ and $f_{100}$ are the fluxes in Jy,
and the far-infrared luminosity $L_{FIR}$ is in solar luminosities.  

	(8) This work. 

	(9) NGC 1316, SH2 and EELR: this work. H$_2$ mass of
NGC 1317 and NGC 1310: Horellou et al. (\cite{horellou95})
after correction for the distance adopted here. 
\end{table*}

\subsection{CO observations}
The observations were done with the 15m Swedish-ESO Submillimeter 
Telescope (SEST)\footnote{The SEST is operated jointly by the European Southern
Observatory (ESO) and the Swedish National Facility for Radio Astronomy, 
Chalmers Centre for Astrophysics and Space Science}
on La Silla, Chile, in two sessions on 1999 December 7--13
and in 2001 January. 
We used the IRAM dual receiver to observe simultaneously at the
frequencies of the $^{12}$CO(1--0) and $^{12}$CO(2--1) lines. 
At 115 GHz the half-power beamwidth of the telescope is 43$''$, which 
corresponds to a linear scale of about 3.9 kpc at the adopted distance of
18.6 Mpc.  
The CO(2--1) map has a resolution of 22$''$ (or $\sim$2 kpc).
This is the highest resolution map of molecular gas in NGC~1316 available
to date. 
A balanced on-off beam-switching mode was used. 
The pointing was regularly checked on nearby SiO masers. 
The temperature scale used throughout this article is the main-beam
temperature $T_{mb}=T_A^*/\eta_{mb}$. 
The main-beam efficiency  of SEST 
is $\eta_{\rm mb} = 0.7$ at 115 GHz and 0.5 at 230 GHz. 
The data were analyzed using the software CLASS. 
The spectra were smoothed to a final velocity resolution 
of $\sim$14.5 \kms.
Only first-order baselines were subtracted. 

\begin{figure*}
\psfig{file=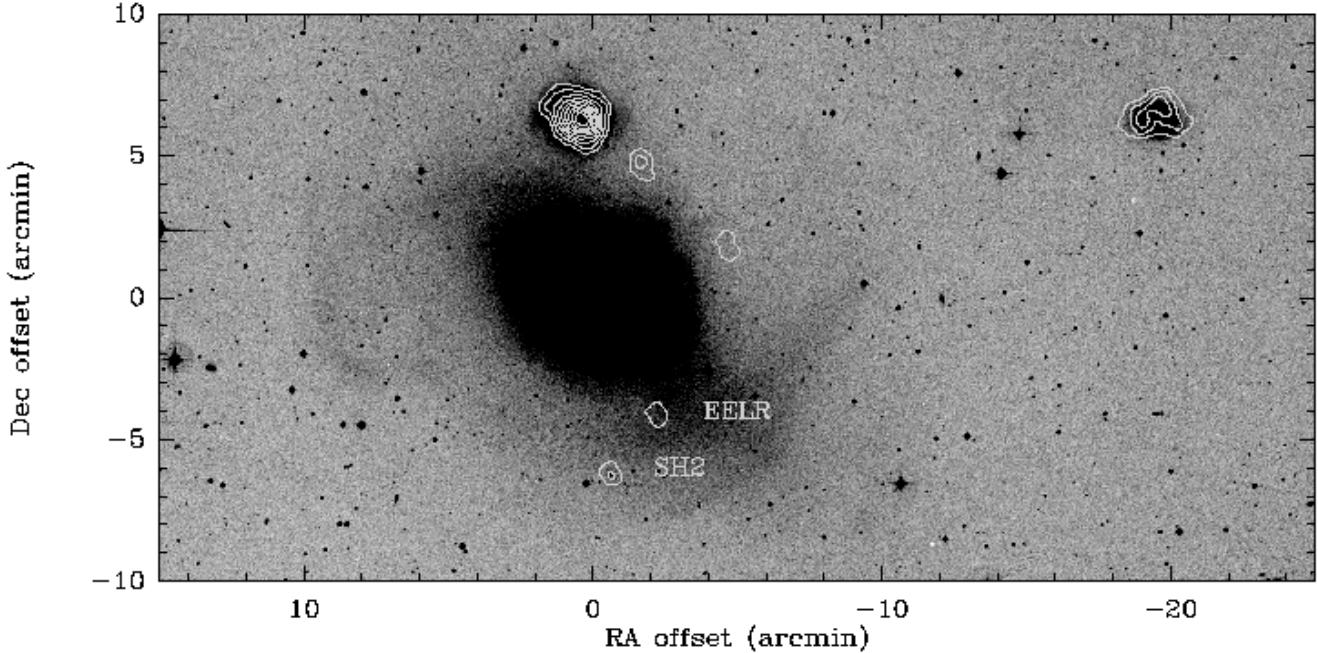,width=18cm}

\caption{
Integrated \hi map overlaid on an optical image from the Digitized Sky Survey. 
The contour levels are $n\times 5.2\times 10^{19}$ cm$^{-2}$. 
The first contour is at the 4$\sigma$ level. 
The two giant \hii regions
to the south, SH2 and EELR, are detected. 
No primary beam correction has been applied to this map.  
}
\end{figure*}

\subsection{Determination of gas masses}  

The integrated \hi masses are calculated using:
$$M_{\rm HI} [M_{\odot}] = 2.356 \times 10^5 D^2_{\rm Mpc} \int S_{\rm HI} dv $$
where $D_{\rm Mpc}$ is the distance in Mpc and $\int S_{\rm HI} dv$ is the integrated
\hi flux in Jy km s$^{-1}$.  

\hi column densities can be derived from the observed integrated intensities by
using the relation: 
$$N_{\rm HI} = {{110.4\times10^3}\over{a b}} \int S_{\rm HI} dv $$
where $N_{\rm HI}$ is the column density in 10$^{19}$ cm$^{-2}$ per beam per channel, 
and $a$ and $b$
are the diameters of the synthesized beam at full width half maximum in arcseconds. 

From the $^{12}$CO(1--0) line intensities we can estimate the mass of molecular 
gas (H$_2$) within the 43$''$ SEST beam:
$$M(H_2) [M_\odot] = 1.25\times 10^5\
                    (\theta /43'')^2\ 
		 	I_{mb}(CO)\
			D_{\rm Mpc}^2
$$
where $\theta$ is the half-power beam width of the telescope and 
$I_{mb}$=$\int T_{mb} dv$ in K~km~s$^{-1}$.   
To be consistent with previous papers, we have used a conversion factor
$X = N(H_2)/I(CO)$ of 2.3 $\times$ 10$^{20}$ 
mol cm$^{-2}$ (K~km~s$^{-1}$)$^{-1}$ 
(Strong et al. \cite{strong88}). 
However, use of the Galactic CO-H$_2$ conversion factor may lead to 
an underestimate of the molecular gas mass for elliptical galaxies 
with a significant amount of cold dust (Wiklind et al. 
\cite{wiklind95}).  

\section{Atomic hydrogen}

Figure~2 shows the \hi field of interest around NGC~1316 superimposed on an 
optical image from the Digitized Sky Survey. 
\hi emission was found in the two companion galaxies NGC~1317 (to the north)
and NGC~1310 (to the west). 
No \hi was detected in the main body of NGC~1316 to a limiting mass of 
$\sim 10^8$ M$_\odot$ but some was 
found at four locations in the outer part of the galaxy.
One of these is associated with the outer H{\sc ii} region SH2. 
The other is coincident with  
the extended emission line region EELR.
The two other \hi concentrations to the west and northwest of NGC~1316  
are located at the edge of an optical plateau between the area of X-ray emission 
as observed by ROSAT (Mackie \& Fabbiano \cite{mackie98}) and the
extended western radio lobe (see Figure~1). 
The mass of each of these \hi concentrations is $\sim 2\times 10^7$ M$_\odot$. 
We will defer discussion of these \hi features and the lack of \hi emission
in the main spheroid of NGC~1316 until later and first discuss the \hi properties
of NGC~1317 and NGC~1310. 

\subsection{The neighbor galaxies NGC~1317 and NGC~1310}

\begin{figure*}
\centering
\psfig{file=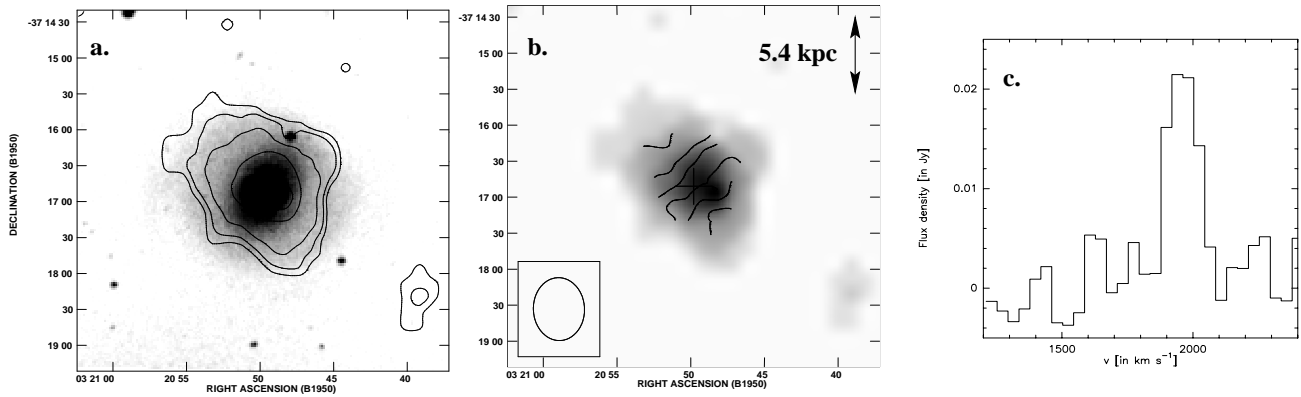,width=18cm,angle=-90}
\caption{
{\bf a.} \hi integrated intensity map of NGC~1317 overlaid on a grey scale 
  optical image from the Digitized Sky Survey. 
A correction for the primary beam has been applied. 
  The levels are  
(4, 8, 16, 31, 47)$\times$10$^{19}$ cm$^{-2}$.
The first contour corresponds to the 2$\sigma$ level. 
{\bf b.} \hi velocity field overlaid on a grey scale image of the
  \hi integrated intensity. 
  Contours of constant heliocentric radial velocity begin at $v= 1910$
  \kms\ in the northeast and increase in steps of 
  20 \kms\ toward the southwest.
  The cross marks the position of the optical center of NGC~1317. 
{\bf c.} Integrated \hi line profile derived from the observations. 
}
\end{figure*}

\begin{figure*}
\centering
\psfig{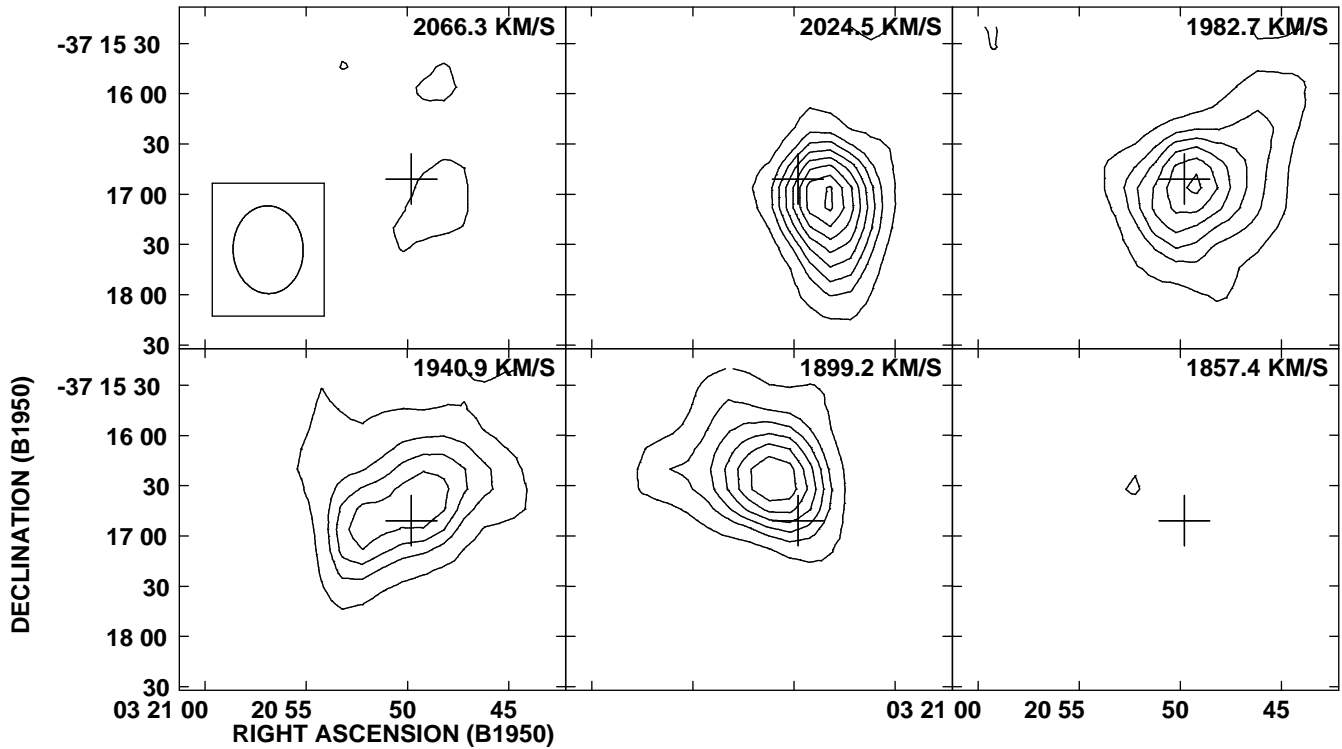}
\caption{The \hi line emission of NGC~1317 displayed as channel maps running from 
2066 \kms\ (top left panel) to 1857 \kms\ (bottom right panel). 
North is up and east is left. 
The levels are 1.5, 3, 4, 5, 6, 7, 8, 9 mJy/beam. 
The first contour corresponds to the 2$\sigma$ level. 
The cross marks the position of the optical center of
NGC~1317. 
}
\end{figure*}

\begin{figure*}
\centering
\psfig{file=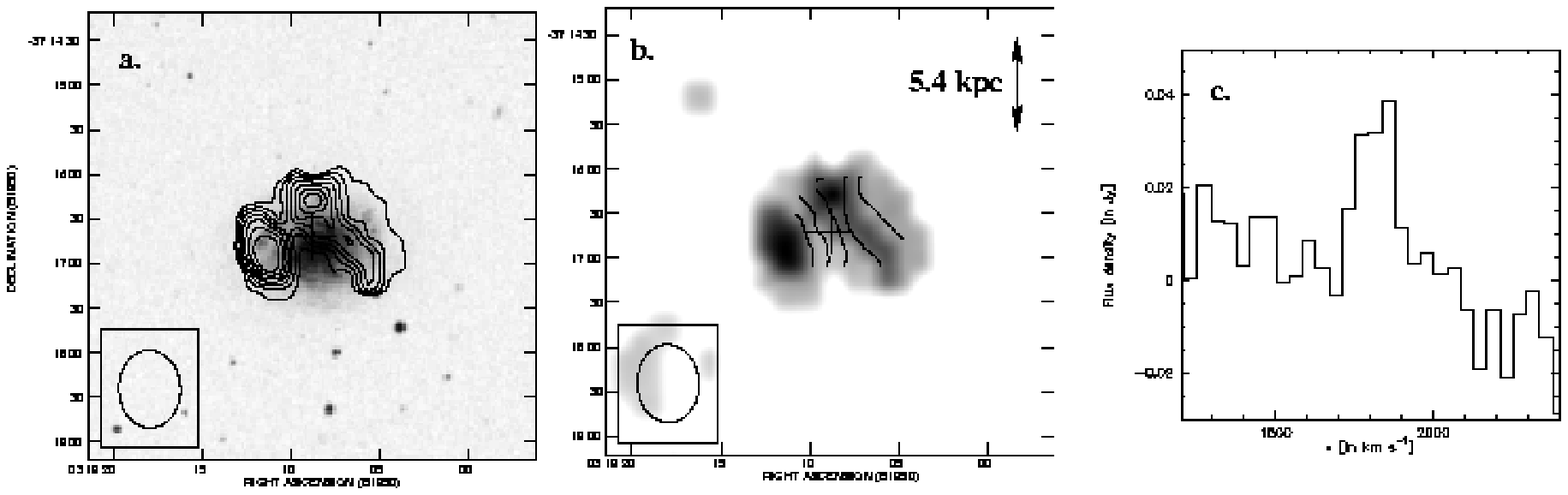,width=18cm}
\caption{
{\bf a.} \hi integrated intensity map of NGC~1310 overlaid on a grey scale
  optical image from the Digitized Sky Survey.
A correction for the primary beam has been applied. 
The levels range from 
21 to 52 $\times$10$^{19}$ cm$^{-2}$ by step of 5.2$\times$10$^{19}$ cm$^{-2}$. 
The first contour corresponds to the 4$\sigma$ level.
{\bf b.} \hi velocity field overlaid on a grey scale image of the
  \hi integrated intensity.
  Contours of constant heliocentric radial velocity begin at $v= 1770$
  \kms\ in the east and increase in steps of
  20 \kms\ toward the west.
  The cross marks the position of the optical center of NGC~1310.
{\bf c.} Integrated \hi line profile derived from the observations.
}
\end{figure*}

\begin{figure*}
\centering
\psfig{file=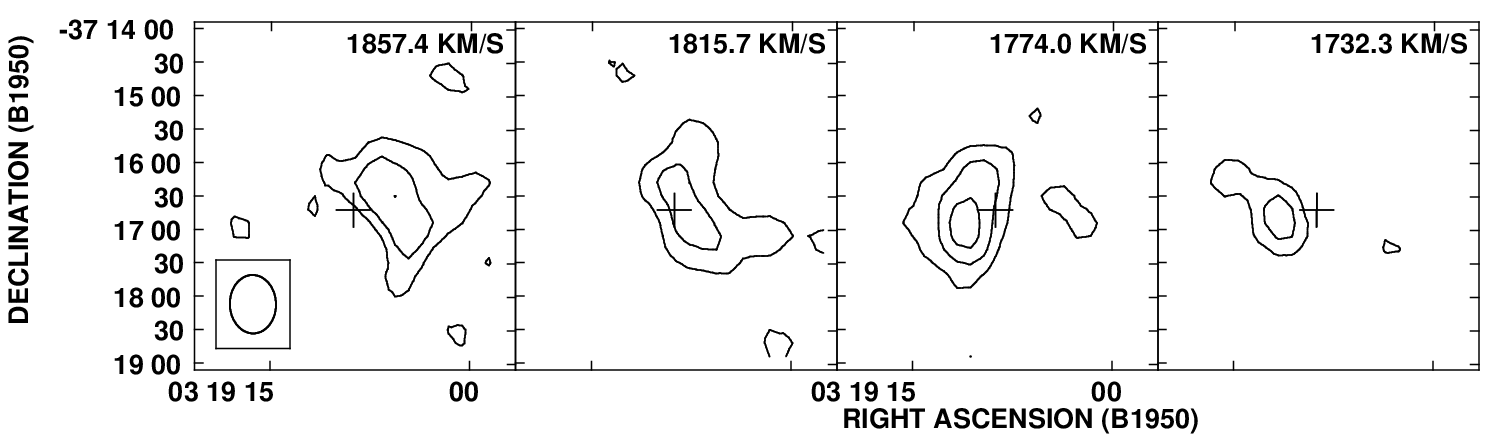,width=18cm,angle=-90}
\caption{The \hi line emission of NGC~1310 displayed as channel maps running from 
1857 \kms\ (left panel) to 1732 \kms\ (right panel). 
North is up and east is left. 
The levels are 4, 8 and 12  mJy/beam. 
The first contour corresponds to the 2$\sigma$ level. 
The cross marks the position of the optical center of
NGC~1310. 
}
\end{figure*}

NGC~1317 is a small barred galaxy north of NGC~1316. 
The two galaxies have a projected separation on the sky of 6$\farcm$3 
(34 kpc) and a radial velocity difference of +165 \kms.  
An inner bar perpendicular to the primary bar is visible in the optical images 
of Schweizer (\cite{schweizer80}) and Wozniak et al. (\cite{wozniak95}).
A ring of star formation is seen in the optical images and in 
the ultraviolet image obtained with the
Ultraviolet Imaging Telescope.  
The primary bar is ``fat" ($\sim$60$''$$\times$80$''$).  
The secondary bar has a position angle of $\sim$60$^\circ$ and
a total length of $\sim$14$''$. 
S80  assigns NGC~1317 a late SB(rs)a or early SB(rs)b type in the 
Hubble classification.  

Figure~3 presents in form of isocontours 
the atomic hydrogen distribution 
and the velocity field in NGC~1317 derived from the observations. 
Also shown is the global \hi spectrum. 
The channel maps are shown in Figure~4. 
The \hi distribution is more extended
than the beam size and is elongated in the direction of the minor axis of the galaxy. 
The gas is clearly rotating, with the south-west side receding. 
The velocity and the linewidth of the global \hi spectrum agree with those of the 
$^{12}$CO(1--0) line detected 
by Horellou et al. (\cite{horellou95}) (see Table~2).  
NGC~1317 has comparable amounts of molecular and atomic gas 
($\sim$3$\times$10$^8$ M$_\odot$). 

The \hi disk of NGC~1317 is small compared to the optical extent. 
The \hi isophote corresponding to a column density of 10$^{20}$ cm$^{-2}$ 
crosses the major axis of the galaxy 
at a radius of about 55$''$  whereas the optical isophote 
at 25 mag arcsec$^{-2}$ is found at 82$\farcs$5. The ratio of the extent of the
\hi disk to that of the optical image is therefore 0.7, which places NGC~1317, if 
classified as an Sab galaxy, 
in the group of galaxies with an \hi distribution strongly affected by the
environment (group {\sc{iii}} of Cayatte et al. \cite{cayatte94} for Virgo galaxies). 
Cayatte et al. (\cite{cayatte94}) showed that those Virgo galaxies with a truncated 
\hi disk have a normal central \hi surface density average as compared to field galaxies 
and concluded that the galaxies (all located close to the cluster core) have been 
affected by ram-pressure
sweeping. NGC~1317 has a central \hi surface density averaged
over half the optical radius of $\sim$3$\times$10$^{20}$ cm$^{-2}$, 
which is lower than that of Sab galaxies in the comparison sample of Cayatte et al.
(5.0$\pm$1.2$\times$10$^{20}$ cm$^{-2}$), but in the range of values found for galaxies 
of earlier types (S0, S0/a, Sa have $3.4\pm0.9$ $\times$ 10$^{20}$ cm$^{-2}$). Given 
the uncertainty in the morphological type and the fact that we are discussing
an individual galaxy, it is difficult to draw any firm conclusion about the central \hi 
density of NGC~1317. Nevertheless, it is clear that the \hi disk is small and that 
NGC~1317 has a low \hi mass for its optical size and morphological type.  We calculate 
an \hi deficiency of .64 in the log, or a factor of 4.4 using the average value for 
isolated Sab galaxies observed by Haynes \& Giovanelli (\cite{hg84}) and the definition: 
${\rm HIdef} = \log (M(HI)/D^2)_{\rm NGC 1317} - \log (M(HI)/D^2)cs $ where $cs$ stands for 
control sample. 
Although NGC~1316 is not located in the center of the Fornax cluster, 
it does dominate its corner of the cluster and is a source of X-ray emission. 
It is possible that NGC~1317
has passed through the X-ray halo of NGC~1316 and been stripped of some of its atomic gas. 

NGC~1310 is an almost face-on spiral fortuitously positioned between us and the western
radio lobe of Fornax~A. Ionized gas from NGC~1310 produces a Faraday screen that 
depolarizes the background radiation (Fomalont et al. \cite{fomalont89}; 
Schulman \& Fomalont \cite{schulman92}). 
Although the galaxy is located 20$\farcm$5 away from the center of Fornax~A
on the plane of the sky  
and in the outer region of the primary beam of the VLA, \hi emission is well detected.  
Figures~5 and 6 present the results of the \hi observations in the same form as those
for NGC~1317.  The approaching gas is located on the eastern side of the galaxy. 
If the spiral arms are trailing, then the northern part of NGC~1310 is the near side. The 
\hi distribution is slightly asymmetric and the emission is stronger on the eastern side. 
We estimate a total mass of atomic hydrogen of 4.8$\times$10$^8$ M$_\odot$ 
for NGC~1310, which is about 20 times higher than the upper limit on the H$_2$ mass
set by Horellou et al. (\cite{horellou95}). 

Comparing the extent of the \hi disk to that of the optical galaxy, one finds that the
isocontour of column density 10$^{20}$ cm$^{-2}$ roughly coincides with the blue 
isophote at the 25th mag arcsec$^{-2}$. This is unusual compared to Sc galaxies in
the field, which normally have an \hi diameter larger than their D$_{25}$ diameter 
by a factor of $1.88\pm 0.08$ (Cayatte et al. \cite{cayatte94}). 
However, NGC~1310 is not significantly \hi-deficient compared to the isolated Sc 
galaxies of Haynes \& Giovanelli (\cite{hg84}) (${\rm HIdef}$=0.17 in the log).  
The small size of the disk ($\sim$2$'$ across) 
together with the location of NGC~1310 in the outer part of 
the primary beam of the VLA make it difficult to say more about its \hi properties.  

\subsection{NGC~1316 (Fornax~A)} 

\begin{figure*}
\centering
\psfig{file=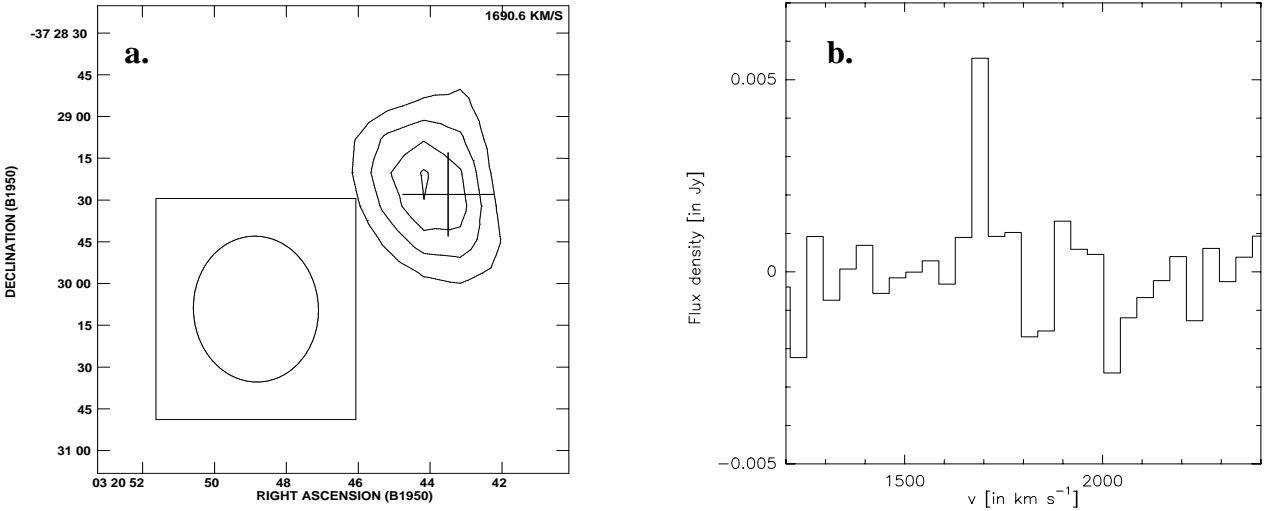,width=18cm,angle=-90}
\caption{
{\bf a.}  
Contour map of the integrated \hi line emission in the \hii region SH2.
The levels are 2, 3, 4 and 5 mJy/beam.
The first contour corresponds to the 2$\sigma$ level. 
The circle shows the size of the beam. 
The cross marks the optical position.
{\bf b.} Integrated \hi spectrum of the \hii region SH2.
}
\end{figure*}

\begin{figure*}
\psfig{file=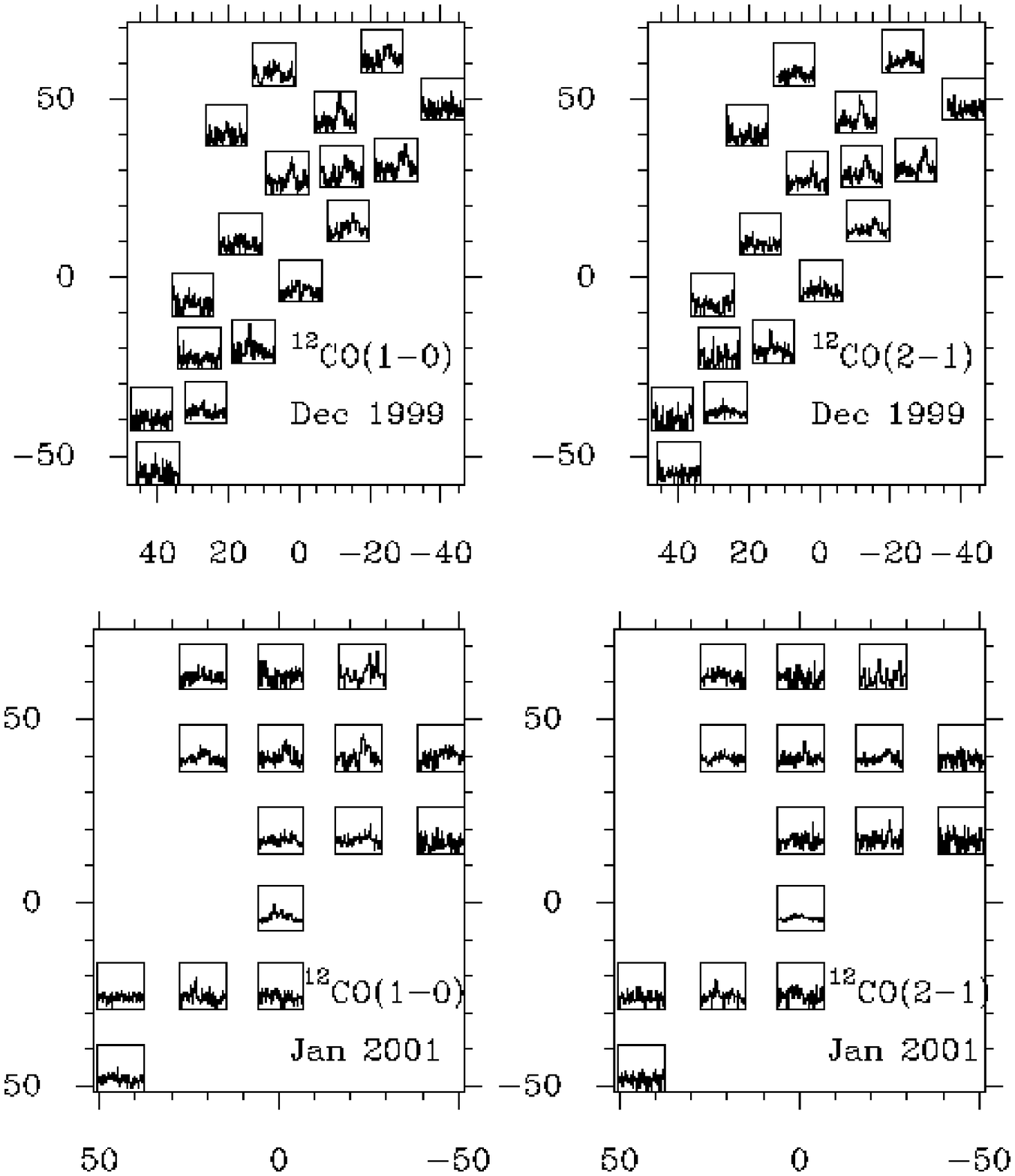,width=16cm}
\caption{{\it Top:} CO(1--0) and CO(2--1) spectra obtained in December 1999. 
The offsets are in arcseconds.  
The $x$ scale ranges from 
1000 to 2300 \kms. The $y$ scale 
(main-beam temperature, $T_{mb}$)
ranges from --0.01 to 0.03 K.
The velocity resolution of the CO(1--0) spectra is 14.5 \kms, 
that of the CO(2--1) spectra is 14.6 \kms.
{\it Bottom:} Results of the January 2001 observing sesion. 
The $x$ scale ranges from 1100 to 2300 \kms. 
The $y$ scale and the velocity resolution are the same as for the spectra above. 
}
\end{figure*}

\noindent{$\bullet$ \it{The spheroid.}}
Huchtmeier \& Richter (\cite{huchtmeier89})
give an upper limit on the \hi integrated line intensity of  4.5 Jy~ km~s$^{-1}$, 
which corresponds
to 3.7$\times$10$^8$ M$_\odot$ at our adopted distance. 
We didn't detect any \hi in the central region of the galaxy down to a 
5 $\sigma$ limit of 10$^7$ M$_\odot$ per beam and per channel. 

If, however, the \hi  emission is as
broad in velocity as the CO detected in the central region (see next section) 
then the
limit is $8.5 \times 10^7$ M$_{\odot}$.  This could be even higher if
the \hi is extended over several beams.  Very extended \hi , moreover, will
be missed in the present observations because the quality of the data at  
the shortest spacings is limited by the presence of 34 Jy of continuum
emission.  A safe limit to the \hi emission in the main body of NGC~1316
is $10^8$ M$_{\odot}$.  

We did not detect \hi absorption towards the nuclear continuum source
which has a flux density of 220 mJy at 1412 MHz.  The 5 $\sigma$ column
density limit is $1 \times 10^{18}$ T$_{\rm s}$ cm$^{-2}$, where
T$_{\rm s}$ is the spin temperature (in K) of the absorbing atomic
hydrogen.

The four \hi detections in the outer regions of NGC~1316 are all far
outside the main body of NGC~1316 and lie at or close to the edge of the
faint optical shells and X-ray emission of NGC~1316.  Each clump
contains 1 to 2 $\times 10^7$ M$_{\odot}$ of \hi.  The southern clumps are
associated with optical emission line gas: the \hii region SH2 
and the EELR.

\noindent$\bullet$ {\it SH2. } 
This {\ion{H}{ii}} region has a size of 
$1\farcs5 \times 3\farcs5$ 
(135 pc $\times$316 pc); 
it is located $6\farcm7$, or 36.2 kpc south of the nucleus and has a radial velocity of
--105$\pm$8 \kms\ relative to it (S80). 
S80 notes that a ``blue fuzz", probably an associated complex of OB stars, 
is visible near the H{\sc ii} region. 

The \hi emission in SH2 appears in one channel centered at 
$v$=1690 \kms\ (our channel width is 41.7 \kms) (see Figure~7). 
This is in agreement with the velocity of 
the ionized gas quoted by Schweizer.  
The \hi mass is about 2$\times$10$^7$M$_\odot$. 
The distribution is unresolved at our resolution of 
$52'' \times 41''$ (4.7 $\times$ 3.7 kpc). 

\noindent$\bullet$ {\it The EELR. } This extended emission line region 
is located $5\farcm3$, or 28.6 kpc from the nucleus and has a size of 
81$''\times27''$,
or 7.3 kpc $\times$ 2.4 kpc (for our distance of 18.6 Mpc). 
On the plane of the sky, it appears on a region of high surface brightness in 
the loop called L$_1$ by S80 and near a major dust lane.  

The \hi associated with the EELR is $\sim 1 \times 10^7$ M$_{\odot}$ and also
unresolved by our beam. 

The two northern clumps are located at the outer edge of a plateau in the optical
surface brightness distribution (Fig.~2, see also Fig.~7 of S80).  
The velocities of the \hi clumps are (from north to south through west)
1857, 2000, 1750 (EELR) and 1690 (SH2), respectively, following the
general rotation pattern on the central CO (next section) and suggesting
that the clumps may be the left-overs of a more widespread \hi structure,
possibly a tidal shred. 
The location and velocity structure of the \hi are reminiscent of 
other shell galaxies such as Centaurus~A (Schiminovich et al.
\cite{schiminovich94})
and NGC~2865 (Schiminovich et al. \cite{schiminovich95}).

\section{Molecular gas}  

\begin{table*}
\def\hf{\hfill}
\caption[]{Line parameters}
\label{table3}
\halign
{#&# 	&#&# 		&# 	&#		&#	&#&#&#&#&#\cr 
\hline
\noalign{\smallskip}
\hf	&\hf		&\hf   &\hf &$^{12}$CO(1--0)&\hf
  	&\hf&\hf				&$^{12}$CO(2--1)	\cr
X&Y 	&RA&DEC 	&$v$ 	&$\Delta v$	&$\int Tdv$    &$\sigma$ 
  	&$v$ 	&$\Delta v$	&$\int Tdv$    &$\sigma$ \cr 
$''$&$''$ &$''$&$''$ &km s$^{-1}$ &km s$^{-1}$&K km s$^{-1}$ &mK
  	&km s$^{-1}$ &{km s$^{-1}$}\hf &K km s$^{-1}$ &mK  \cr 
(1) 	&(2)		&(3)	&(4)		&(5)
   	&(6)	&(7)	&(8)	&(9)\cr
\noalign{\smallskip}
\hline
\noalign{\smallskip}
	&	&	&	&	&	&1999 observing &session\cr
\noalign{\smallskip}
\hline
\noalign{\smallskip}
0	&$+$22	&$+$13.5	&$-$17.3	&1531$\pm$13	&129$\pm$58	&2.0$\pm$0.6	&5.9
				&1550$\pm$6	&88$\pm$13	&1.9$\pm$0.3	&4.9\cr
0	&$+$44	&$+$27.1	&$-$34.7	&1557$\pm$14	&128$\pm$27	&1.1$\pm$0.2	&3.5
				&1525$\pm$7	&75$\pm$17	&0.9$\pm$0.2	&3.0\cr
$-$22	&$-$44 	&$-$9.8	&$+$48.2	&1809$\pm$8	&167$\pm$20	&4.3$\pm$0.4	&5.0
				&1818$\pm$6	&136$\pm$14	&3.3$\pm$0.3	&4.4\cr
0	&$-$44  	&$-$27.1	&$+$34.7	&1914$\pm$16	&263$\pm$31	&4.5$\pm$0.5	&5.6
				&1919$\pm$11	&178$\pm$31	&3.0$\pm$0.4	&4.7\cr
0	&0	&0	&0	&\hf		&\hf		&$<$2.4		&5.1 
			 	&\hf            &\hf            &$<$2.4        &5.0\cr
0	&$-$22	&$-$13.5	&$+$17.3	&1813$\pm$16    &268$\pm$42     &3.4$\pm$0.4  &4.4
				&1859$\pm$11	&186$\pm$24	&2.0$\pm$0.2	&3.0\cr
$-$22	&$-$22 	&$+$3.8	&$+$30.9 	&1778$\pm$9     &178$\pm$21     &3.5$\pm$0.4  &4.8
				&1797$\pm$5	&67$\pm$14	&1.4$\pm$0.2 	&4.3\cr
$-$22	&0	&$+$17.3	&$+$13.5	&\hf       	&\hf       	&$<$2.4       	&5.1
				&\hf            &\hf            &$<$2.1        &4.2\cr
$-$22	&$-$66 	&$-$23.3	&$+$65.6	&1835$\pm$17    &224$\pm$38     &3.4$\pm$0.5  &5.8
				&\hf            &\hf            &$<$2.1        &4.7\cr
$-$22	&$+$22	&$+$30.9	&$-$3.8	&\hf       	&\hf       	&$<$3.3       	&7.1
				&\hf            &\hf            &$<$2.4        &5.4\cr
$-$44	&$-$44	&$+$7.6	&$+$61.8	&\hf       	&\hf       	&$<$2.7       	&5.9
				&\hf            &\hf            &$<$2.1        &4.2\cr
0	&$-$66	&$-$40.6	&$+$52	&\hf       	&\hf       	&$<$2.7       	&5.9
				&\hf            &\hf            &$<$2.4        &5.4\cr
$-$44	&$-$22	&$+$21.1	&$+$44.4	&\hf       	&\hf       	&$<$2.7       	&6.1
				&\hf            &\hf            &$<$3.6        &7.6\cr
0	&$+$66	&$+$40.6	&$-$52	&\hf       	&\hf       	&$<$3.3       	&7.3
				&\hf            &\hf            &$<$2.4        &5.0\cr
$-$11	&$-$33	&$-$11.6	&$+$32.8	&1807$\pm$19    &206$\pm$48     &3.5$\pm$0.7  &7.7
				&1802$\pm$11	&202$\pm$32	&3.5$\pm$0.4	&4.6\cr
$-$11	&$+$55	&$+$42.5	&$-$36.6	&\hf       	&\hf       	&$<$2.7        &5.9
				&\hf            &\hf            &$<$4.8        &10.6\cr
$-$11	&$+$33	&$+$29.0	&$-$19.2	&\hf      	&\hf       	&$<$2.4        &5.0
				&\hf            &\hf            &$<$3.3       &7.0\cr
\noalign{\smallskip}
\hline
\noalign{\smallskip}
	&	&	&	&	&	&2001 observing &session\cr
\noalign{\smallskip}
\hline
\noalign{\smallskip}
   0.0&    0.0&     0&     0&    1634	&343	&3.4  &2.1
		        &1653    &208     &2.0 &2.1 \cr
 $-$27.7&   $-$35.5&     0&    $+$45&    1829$\pm$9   &157$\pm$22   &3.2$\pm$0.4 &5.4
			  &1795$\pm$4  &48$\pm$9 &1.7$\pm$0.3  &3.4 \cr
 $-$10.4&   $-$49.0&   $-$22&   $+$45&   1846$\pm$12  &190$\pm$25  &4.3$\pm$0.5 &6.4
			     &1910$\pm$13 &186$\pm$33  &2.9$\pm$0.4 &5.4 \cr
 $+$7.8&   $-$63.2&   $-$45&    $+$45&    1900$\pm$35 &400$\pm$70  &4.7$\pm$0.8 &4.6
			&\hf            &\hf            &$<$1.5        &9\cr  
 $-$45.0&   $-$21.9&    $+$22&    $+$45&   1737$\pm$14 &242$\pm$27  &2.6$\pm$0.3 &3.4   
			&1697$\pm$27 &270$\pm$53  &2.6$\pm$0.3 &4.8 \cr
 $-$59.2&   $-$40.0&   $+$22&   $+$68&    1671$\pm$30 &207$\pm$53   &1.4$\pm$0.4 &4.8    
			   &1578$\pm$52 &394$\pm$96  &3.3$\pm$0.9 &6.8\cr
   $+$3.8&   $-$30.9&   $-$22&    $+$22&   1898$\pm$22 &292$\pm$60 &2.7$\pm$0.4 &4   
			 &1967$\pm$5 &41$\pm$10  &1.6$\pm$0.4 &9\cr
  $-$3.8&    $+$30.9&    $+$22&   $-$22&    1514$\pm$6 &73$\pm$13   &1.5$\pm$0.3 &5.3   
			  &1529$\pm$7 &103$\pm$19  &2.8$\pm$0.4  &7\cr
 $-$13.5&    $-$17.3&     0&    $+$22&    1805$\pm$41 &333$\pm$75  &1.1$\pm$0.4 &4.2
			&\hf            &\hf            &$<$1.8        &7.8\cr    
  $+$21.9&   $-$45.0&   $-$45&    $+$22    &\hf      	&\hf       	&$<$1.8        &7.1
			&\hf            &\hf            &$<$4.5        &12.\cr    
  $+$13.5&   $+$17.3&     0&  $-$22&     1506$\pm$21 &142$\pm$44  &1.1$\pm$0.3 &4.7  
			  &1500$\pm$36  &327$\pm$74  &2.7$\pm$1.0 &10. \cr
 $-$21.9&    $+$45.0&    $+$45&   $-$22     &\hf      	&\hf       	&$<$0.9        &4.
			&\hf            &\hf            &$<$2.1        &8.\cr     
  $-$7.8&    $+$63.2&    $+$45&   $-$45      &\hf      	&\hf       	&$<$1.2        &4.4
			&\hf            &\hf            &$<$2.4        &9.\cr     
 $-$41.9&   $-$53.6&     0&    $+$68     &\hf      	&\hf       	&$<$1.8        &5.4
			&\hf            &\hf            &$<$4.5        &12.\cr     
 $-$23.7&   $-$67.7&   $-$23&    $+$68&     1963$\pm$29 &334$\pm$47  &3.6$\pm$0.8 &8.7  
			 &1960$\pm$32 &72$\pm$56 &1.2$\pm$1.0  &15. \cr
\noalign{\smallskip}
\hline
\noalign{\smallskip}
	&	&	& 	&Combined &data and &outer H{\sc ii} regions\cr
\noalign{\smallskip}
\hline
\noalign{\smallskip}
   0.0	&    0.0&     0	&     0	&1559	&189	&2.4  	&2.1
		        	&1642 	&213	&1.9	&2.1 \cr
   EELR	& 	&	&	&		&		&		&3.2 
				&		&		&		&3.8\cr
   SH2&		&	&	&		&	&		&3.1 		&	
				&		& 	&4.1\cr
\noalign{\smallskip}
\hline
\noalign{\smallskip}
}
\begin{list}{}{}
\item[(1),](2): Offset positions after rotation of the 
coordinate system by PA=142$^{\circ}$.
\item[(3),](4): Offset positions in RA and DEC. 
\item[(5)] to $(7)$: 
For all positions but the center we give the 
parameters of the gaussian lines fitted to the CO(1--0) spectra: 
central velocity, linewidth, integrated main-beam intensity. 
For the center position where the lines are clearly not gaussian, 
we give the velocity of the centroid, the linewidth and 
the main-beam intensity integrated over the velocity range
of the line (see Figure~9). 
For non-detections, the limit on the integrated intensity was taken as
3$\times \sigma\times 150$ km s$^{-1}$.
\item[(8):] Noise per 14.5 km s$^{-1}$ channel. 
\item[(9)] to (12): 
Same as columns (5) to (8) but for the CO(2--1) spectra. 
\end{list}
\end{table*}

\subsection{The central region of NGC~1316 (Fornax~A)}

Results from the CO observations are presented in Figures~8 to 10 
and the line parameters are given in Table~3. 
The observations were done by two observers in two separate
sessions, and the observing strategy was slightly different. 
During the first observing session 
(see top panel of Fig.~8), 
we mapped the CO emission along the axis with a position angle PA=142$^{\circ}$,
given by S80 as the direction of the maximum extent of the ionized gas 
and probably the rotation axis of the spheroid
(perpendicular to the optical major axis). 
During the second session (Fig.~8, bottom panel), the grid was in RA/DEC. 
Emission is detected north-west and south-east of the nucleus as well as 
toward the center.
The intensities in the NW region are about twice as strong as in the SE, 
both in CO(1--0) and CO(2--1).  
The CO distribution roughly follows that of the dust along the minor
axis and to the NE.

$\bullet$ {\it Emission was detected toward the central position.} 
The averaged central spectra obtained during both observing sessions 
are shown in Figure~9. 
The noise level in our CO(1--0) spectrum is about 5 times lower than that in the
spectrum obtained by 
Wiklind \& Henkel (\cite{wiklind89}) and the bandwidth is larger, which makes it
easier to see the line structure. The integrated line intensity we infer is about three
times lower than that reported in that work, but it may agree within the not quoted 
error bars of Wiklind \& Henkel. 
Sage \& Galletta (\cite{sage93}) had detected CO(1--0) at two positions 30$''$ to the SE and
50$''$ to the NW of the nucleus but not toward the center. The fluxes and the limits
quoted agree with ours within the uncertainties.  
The CO(1--0) line seems to be very broad ($\sim$500 km s$^{-1}$ at the base) 
and to consist of three
components. It is likely that the molecular gas distribution is discrete and that
the low-velocity component arises from emission from the SE that enters
the 43$''$ beam,
whereas the hint of emission at high velocity ($\sim$1900 km s$^{-1}$) comes from the 
NW concentration. All those features are, if present at all, much weaker in the 
CO(2--1) spectrum taken with a narrower beam.   
We calculate a mass of the H$_2$ gas of 
$\sim$10$^8$  M$_\odot$ within the inner 43$''$ (3.9 kpc). 
This estimate of mass is based upon a standard Galactic conversion factor.

$\bullet$ {\it The NW concentration} is centered at   
$\Delta \alpha$=--14$''$, $\Delta \delta$=+37$''$ relative to the central position,
that is, at a projected distance of 3.5 kpc from the nucleus. 
The emission is extended on the scale of the 22$''$ CO(2--1) beam, 
and the measured apparent diameter at half maximum intensity 
is 35$''$, which 
correspond to a true extent of $\theta_{\rm source}$ 
= 27$''$, or a linear size of $\sim$2.5 kpc. 
This is consistent with the CO(1--0) measurements, which yield an apparent diameter 
of the half peak intensity of 50$''$ (and a true source size of 24$''$).  
Velocities around 1850 \kms\ are found in the NW region, 
that is, $\sim$70 km s$^{-1}$ higher than the systemic velocity 
of NGC~1316. 
A slight east-west velocity gradient is observed, with higher velocities to the 
west. 

From the apparent extent of the northwest CO concentration measured above and the 
CO(1--0) peak intensity, we derive a mass of molecular hydrogen within that
region of 2.2$\times$10$^8$ M$_\odot$.

$\bullet$ {\it In the southeast}, the strongest emission is detected 22$''$ 
($\sim$2  kpc) away from the nucleus and is half as strong at 44$''$. 
It corresponds to a H$_2$ mass 
of 0.86$\times$10$^8$ M$_\odot$.
The velocities measured there are of the order
of 1550 \kms, that is, $\sim$230 \kms\ lower than the systemic velocity
of NGC~1316.

$\bullet$ {\it The CO velocity fields} are displayed in Fig.~10c and 10d. They  
are fairly regular. The systemic velocity of NGC~1316, however, is likely to be higher
than suggested from those interpolated contour maps. As discussed above 
for the central position, 
the low velocity wing around 1500 \kms\ could be due to emission 
from the NW 
concentration that enters the beam. 
The centroid velocity of 1559 \kms\ given in Table~2 for 
the central position would therefore be too low and affect the 
interpolation in the contour map. 

\begin{figure}
\centering
\psfig{file=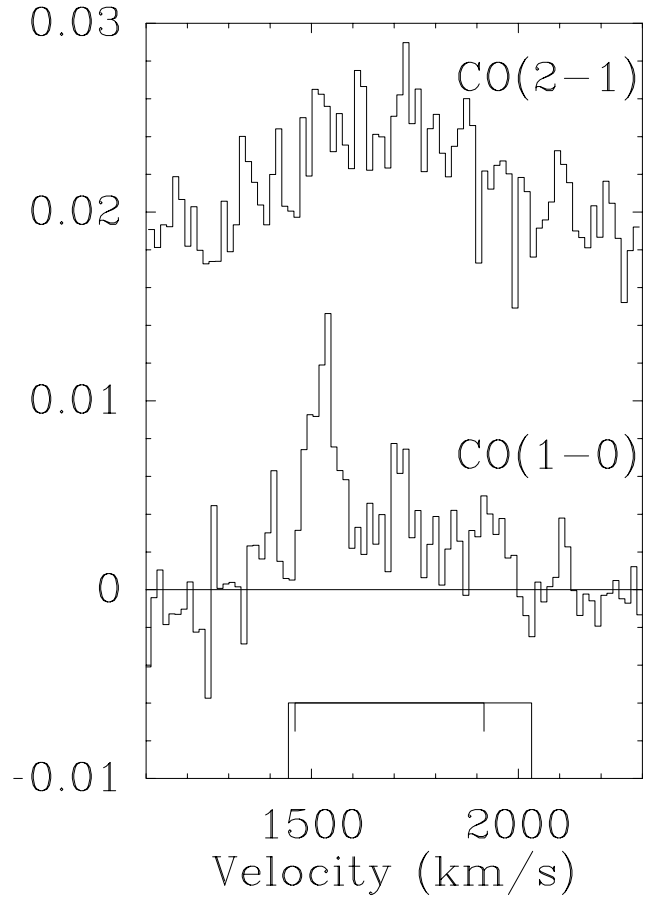,width=8.8cm}
\caption{CO(1--0) detection toward the center of NGC~1316. 
The line is broad and seems to consist of three components. 
The $y$ axis shows the main-beam temperature in Kelvin. 
The width of the window set to subtract a baseline is shown
on the $x$ axis: 1444 to 2032 \kms\ for the CO(1--0) spectrum
as indicated by the box, and 1460 to 1917 \kms\ for the CO(2--1)
spectrum as indicated by the ticks.
The CO(2--1) line seems to be narrower than the CO(1--0) one, 
which is to be expected if the molecular gas is located in the
dusty regions offset from the center and located outside the 
narrower CO(2--1) beam. 
The spectra have been produced by averaging the data taken during 
both observing sessions. 
}
\end{figure}

\subsection{The outer emission-line regions}   
We have searched for CO emission toward the giant H{\sc ii} region SH2 
and the extended emission line region  
but have not detected any,
down to the noise levels quoted in Table~3. For SH2 as for EELR, this corresponds 
to a 5$\sigma$ limit on the H$_2$ mass of $\sim$10$^7$ M$_\odot$  per 43$''$ beam per 
14.5 \kms\ channel. 

\section{Discussion}

While CO is clearly detected in the central region of NGC~1316,
atomic hydrogen is not. 
The distribution of the molecular gas follows that of the dust,  
and its kinematics provides new
insight into the dynamical history of the system. 
\hi is detected in the outer part where no CO is found down to a 
similar mass limit. 

\subsection{Comparison with other components of the ISM}

\begin{figure*}
\centering
\psfig{file=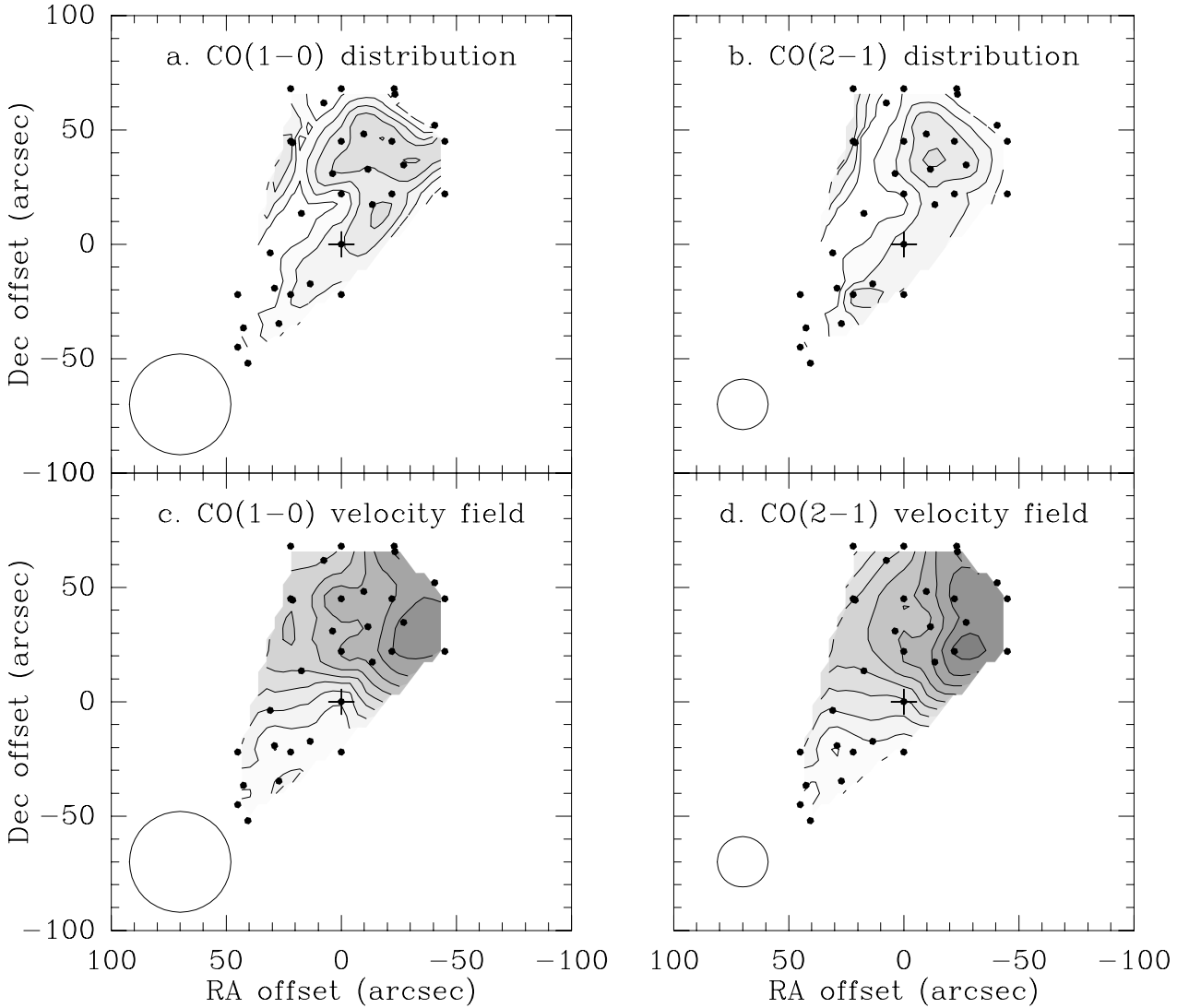,width=22cm,angle=-90}
\caption{{\bf a. and c.} Contour map of the CO(1--0) distribution and 
corresponding velocity field. 
The levels of the total intensity map go from 0.5 to 6 K~km~s$^{-1}$ in steps
of 1 K~km~s$^{-1}$. 
The levels of the velocity map go from 1500 to 2000 \kms\ in steps of 
50 \kms. 
The circle shows the size of the beam. 
The black dots indicate the observed positions. The central position is marked by a cross. 
{\bf b. and d.}  Contour map of the CO(2--1) distribution and
corresponding velocity field. 
The levels are the same as for the CO(1--0).
}
\end{figure*}

$\bullet$ {\it The central part of NGC~1316} 

The dust patches are very visible in the $B-I$ {\it HST} image of NGC~1316 
displayed in Figure~11c. 
They appear on the NW and the SE side of the nucleus, roughly aligned along the
minor axis of the galaxy. The SE lane can be traced 
to within 1 arcsec from the center, which led Schweizer (\cite{schweizer80}) 
to suggest that
it lies on the near side of the galaxy.   
The molecular gas distribution at 22$''$ resolution follows roughly the distribution of 
the dust. 

Ionized gas 
was found by S80 to be associated with the dust and distributed in a rotating disk 
with extremal velocities $\pm 350$ \kms\ at 84$''$ NW and 35$''$ SE. 
The SE side of the disk is approaching.  The extreme range of CO velocities that we 
measure on either side of the nucleus, 1500 to 1960 \kms\ 
at 32$''$ SE and 72$''$ NW, is thus smaller
than the range of velocities in the ionized disk.

Shaya et al. (\cite{shaya96}) used {\it HST}
$V$ and $I$ band images of the central region 
(a field of view of 70$''$) to analyze the three-dimensional 
distribution of the dust. They derived a small amount of extinction 
toward the very center, $A_V<$ 0.4 mag and $A_I<$ 0.2 mag, and
$A_V\sim$1.5 in the regions of maximum extinction. 
The amount of molecular gas that we infer from the CO observations
correlates with the amount of extinction: we detect only $\sim$10$^8$
M$_\odot$ in the center and about twice as much in the dusty NW region. 

The radio jet observed by 
Geldzahler \& Fomalont (\cite{geldzahler84}) 
could be deflected by interstellar gas, especially in the NW.  Figure 11a
shows a superposition of our CO(2--1) map and the radio jet at $\lambda$20 cm.
The direction of the central jet changes abruptly in the NW just below the
peak in the CO distribution. There is a hint of a similar bending near the 
SE peak in the CO. 

The most luminous galaxies in the far-infrared are mergers.  
NGC~1316 was detected by IRAS in the four bands but its far-infrared luminosity
is moderate ($\sim$2$\times$10$^9$ L$_\odot$). 
The $L_{FIR}$ to $M$(H$_2$) ratio, often used as an indicator of the efficiency
at which molecular gas is converted into new stars, is $\sim$5. 
This is lower than the average value of $\sim$15 
that we calculated for
the sample of ellipticals
observed by Wiklind et al. (\cite{wiklind95})  
(after exclusion
of the two galaxies with the highest values of $L_{FIR}/M({\rm H}_2)$ ratios, 
an S0 and a galaxy larger than the beam size, for which the H$_2$ mass must have been
underestimated). 
Nevertheless, this may not mean that the star formation efficiency of 
NGC~1316 is 
lower 
than in other ellipticals since it is known that some of the
far-infrared emission is not related to massive ionizing stars, but to dust
heated by the overall stellar radiation field.  
Moreover, the Galactic CO-to-H$_2$ conversion factor may not apply to ellipticals
or it may vary from one elliptical to another. For example, an elliptical galaxy 
that has accreted a gas-rich
but metal-poor companion could be CO-poor and its H$_2$ mass would be underestimated.   

\noindent$\bullet$ {\it The outer \hii regions} 

It is interesting that SH2, despite being so much smaller than EELR, 
possesses about twice as much atomic gas. 
Mackie \& Fabbiano (\cite{mackie98}) give an H$\alpha$ luminosity 
$L_{\rm H\alpha}$= 5.46$\times$10$^5$ L$_\odot$ 
for the EELR, and a mass of ionized gas
$M_{\rm HII}$=5$\times$10$^3$ M$_\odot$  
(when corrected for our adopted distance). 
If the EELR were photoionized by hot stars, this quantity of
ionized gas could be maintained by $\sim 40$ O-type stars. However,
Mackie and Fabbiano argued that there is no evidence of a population of
young stars associated with the EELR and suggested that the ionized
gas is produced by shocks. 
Although Mackie \&\ Fabbiano
do not quote an ionized gas mass for SH2, it appears from their H$\alpha$
image, which has been smoothed to a resolution of $4{\farcs}3$, that the
H$\alpha$ flux is of the order of $3\times 10^{-16}$ ergs s$^{-1}$
cm$^{-2}$, integrated over the $1{\farcs}5\times 3{\farcs}5$ size reported
by S80. At our adopted distance, such recombination emission can be maintained
by a single O-type star. For comparison, the giant H{\sc ii} region NGC 604 in
the Local Group galaxy M 33 contains $3\times 10^6$ M$_{\odot}$ of ionized gas
(Churchwell \&\ Goss \cite{churchwell99}) and approximately 200 O-type
stars (Gonz\'alez Delgado \&\ P\'erez \cite{gonzalez00}). Although the
H$\alpha$ emission of SH2 may signify recent star-forming activity, the
population of massive stars is not large in comparison with the mass of
neutral gas in the vicinity, $M_{\ion{H}{i}}\approx 2\times 10^7$ M$_{\odot}$.

\subsection{The absence of \ion{H}{i} emission and absorption in the central
spheroid}

In the central region of NGC~1316 the \hi $\lambda$21 cm line is not clearly
detected in emission nor is there any noticeable absorption against the
continuum emission. Our limit on \hi emission in the spheroid of NGC~1316 
could represent a mass of H as high as $10^8$ M$_{\sun}$. The limit on
absorption against the central continuum source corresponds to a
column density of H  $<1\times 10^{18} T_{\rm s}$ cm$^{-2}$,
implying that only gas with spin temperature $T_{\rm s}< 100$ K 
could reasonably have been detected. The upper limit on \hi emission
may at first sight seem surprising in view of the 
large amounts of dust and molecular gas in the spheroid. Fornax~A
is not unique in this respect: other well known elliptical galaxies
and merger remnants show CO and dust 
but no detectable H in their centers
(e.g. NGC~7252, Wang et al. \cite{wang92}; Hibbard et al. \cite{hibbard94}).
In contrast there are smaller elliptical galaxies or merger
remnants that have dust, CO and plenty of H all the way in to their centers.
Examples are NGC~5128, NGC~3656, and Arp~230, all of which have disks
that are viewed nearly edge-on
(Schiminovich et al. \cite{schiminovich94};
Eckart et al. \cite{eckart90}; 
Balcells et al. \cite{balcells01}; 
Sofue et al. \cite{sofue93}; 
Schiminovich et al. \cite{schiminovich01};
Galletta et al. \cite{galletta97}). 
The H-deficient and H-rich systems also
differ in the size of the spheroid and presence of extended X-ray emission.  
However, the detectability of \hi absorption may also depend on whether 
a gaseous disk is present and if so how it is inclined to the observer's
line of sight.

The CO emission from NGC~1316 probably arises in the surfaces of  
molecular clouds. Any neutral atomic gas associated with such 
molecular clouds may cover too little of the central continuum source to
be detectable in absorption in the $\lambda$21 cm line
at the angular resolution and sensitivity of the available measurements.
The high pressure of the hot X-ray-emitting gas and the shocks associated
with the merger events may help confine the neutral gas in small clouds, 
while the X-radiation keeps the more dilute gas ionized. 

Finally, it is interesting to consider briefly how the spin temperature
of H in the central region of NGC~1316 might be affected by the extreme
environment there. There are radiative processes that can boost the
spin temperature of low-density H: absorption and stimulated emission 
induced by the continuum radiation in the 21 cm line itself (cf. Bahcall 
\& Ekers \cite{bahcall69}), and absorption and fluorescence in the 
\hi L$\alpha$ line (Deguchi \&\ Watson \cite{deguchi85}). These processes 
will not affect the interpretation of optically thin \hi emission lines, 
but they could make absorption even less readily detectable.
Although Fornax~A is a bright continuum source at $\lambda$21 cm, the 
integrated flux density of 0.24 Jy over the central $10''\times 30''$ 
(Ekers et al. 1983, Geldzahler \&\ Fomalont 1984) corresponds to an 
average continuum brightness temperature of 565 K and yields an
absorption rate $F=0\to 1$ of $\rho\approx 7\times 10^{-11}$ s$^{-1}$.
The brightness temperature in the extended central jets is lower,
$\sim 150$ K, which might be a lower limit on the spin temperature of
dilute gas in the center. 
Pumping in the \hi L$\alpha$ line will have a larger effect on the 
spin temperature of \hi.  The average surface brightness of the 
\hi L$\alpha$ line can be estimated from the observed surface brightness
in the H$\alpha$ line, $I\sim 4\times 10^{-17}$ erg s$^{-1}$ cm$^{-2}$
arcsec$^{-2}$ (Kim et al. 1998), with the assumption that a fraction $\eta$
of the Balmer-$\alpha$ emissions are followed by emission in L$\alpha$.
The corresponding pumping rate $F=0\to 1$ is $\sim 10^{-7}\eta$ s$^{-1}$,
averaged over the central arcmin of the galaxy. Since $\eta$ is likely
to be of the order of unity, this excitation rate is much higher  
than the mean rate of continuum pumping in the 21 cm line and is
high enough to dominate over spin-changing collisions if the 
density of the \hi-emitting gas is $\la 10^2$ cm$^{-3}$.  When
L$\alpha$ pumping controls the spin temperature, this temperature
is probably $\sim 10^4$ K (Deguchi \&\ Watson \cite{deguchi85}). Thus
the L$\alpha$ pumping could make the \hi line undetectable in absorption in
the center of NGC~1316 by ensuring that its spin temperature remains high 
even where the kinetic temperature might be much lower than $10^4$ K.
The true magnitude of the L$\alpha$ pumping effect is uncertain, of course,
because the distributions of dilute neutral gas and of L$\alpha$ photons
are not known.

There is a large amount of hot, ionized, X-ray-emitting gas in the central
kpc of NGC~1316, which arises at the expense of neutral gas that might
otherwise exist there. In the future, it will be valuable to obtain
\hi observations of greater sensitivity and higher resolution to try to
trace the neutral gas and its interactions with its environment in more
detail.

\subsection{Timescales}

\begin{figure*}
\centering
\psfig{file=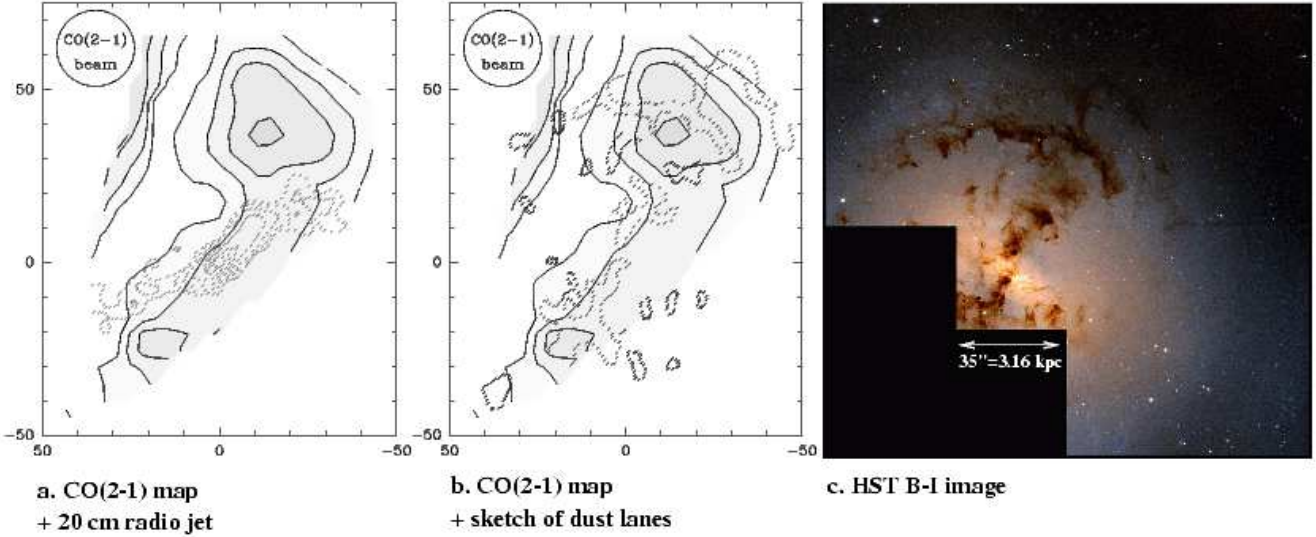,width=18cm}
\caption{
{\bf a.} CO(2--1) contour map overlaid on 
the 20 cm radio jet mapped by Geldzahler \& Fomalont (\cite{geldzahler84}). 
{\bf b.} CO(2--1) contour map overlaid on 
Schweizer (1980)'s sketch of the dust lanes (dots). 
{\bf c.} {\it HST} B-V image of the central region of NGC~1316 
(press release STScI-PR99-06,  Grillmair et al. \cite{grillmair99}).  
}
\end{figure*}

The complex structure of NGC~1316 (five tails or loops of varying morphology) 
argues for a rich dynamical history and probably a succession of merging events. 
The presence of shells around ellipticals and minor-axis dusty disks 
can be interpreted in two different ways: as tracers of the accretion of 
a small companion onto an already existing elliptical galaxy (e.g. Quinn \cite{quinn84}), 
or as the returning of tidal material during the merger of two disk galaxies
(e.g., Hernquist \& Spergel \cite{hernquist92}, 
Balcells \cite{balcells97}).
There is observational evidence that NGC~1316 formed in a major merger $\sim$3 Gyr ago 
and has been accreting material more recently, as discussed below.

\subsubsection{A major merger $\sim$3 Gyr ago}

NGC~1316 has a relaxed morphological appearance within 3$'$, 
which indicates that, if the galaxy is the result of a major merger between two disk
galaxies, the event 
must have occurred more than $\sim$2$\times 10^8$
years ago. 

The population of globular clusters constrains the age of the system. 
Among 24 star clusters associated with NGC~1316, Goudfrooij et al. (\cite{goudfrooij01})
identified four exceptionally luminous ones ($M_V < 12.3$) with near-solar metallicities,
whose age of 3.0 $\pm$ 0.5 Gyr 
could be determined from comparison with single-burst population models.
Those intermediate-age clusters would have formed during a merger of two disk galaxies 
that resulted in the formation of the elliptical whereas the older ($\sim$14 Gyr old)
metal-poor clusters are likely to have belonged to the progenitor disk galaxies. 
As a remnant of the merger of two disk galaxies, NGC~1316 would be at a more advanced
stage than for instance NCC 7252, which is believed to be 0.5--1 Gyr old and still
displays extended tidal tails (Schweizer \& Seitzer 
\cite{schweizer98}).  

Optical spectra of the inner 3$''$ of NGC~1316 have revealed a stellar population
of age $\sim$2 Gyr (Kuntschner \cite{kuntschner00}). Those stars would therefore have
formed from the molecular gas that fell toward the center following the merger event
and would have contributed to the depletion of molecular gas observed in the central 
region today. Strong nuclear H$\beta$ absorption has been found  
(Schweizer \cite{schweizer81}; Bosma et al. \cite{bosma85}). 
It might indicate an important component of A-type stars which are part of a 
post-starburst population.  

The X-ray luminosity increases as a result of gas being shock heated or evaporated by
electrons in the already existing hot plasma 
(e.g., Goudfrooij \& Trinchieri \cite{goudfrooij98})
and mass loss from evolving giant stars and supernovae Ia 
(e.g., Brighenti \& Mathews \cite{brighenti99}).  
Using ROSAT, Kim et al. ({\cite{kim98}) 
detected $\sim 5\times 10^8$ M$_\odot$ (at $D=18.6$ Mpc) of hot 
($\sim$10$^7$ K) gas, which is small compared to the amount of hot gas found in some 
early-type galaxies of comparable optical luminosity. However, 
Kim et al. (\cite{kim98}) point out that the amount of hot gas 
is comparable with what would be expected from stellar evolution in
$\sim$1 Gyr. A comparable amount of hot gas 
is coincident with the tidal tails in the outskirts
of NGC~1316 (Mackie \& Fabbiano \cite{mackie98}) and may be due to more recent events. 

\subsubsection{More recent infalls}

The presence of loops and ripples as well as of a disk of ionized gas 
is strong evidence in favor of the infall of a small galaxy onto NGC~1316
less than 1 Gyr ago (Schweizer \cite{schweizer80}).  
In particular, the giant loop called L$_1$ by Schweizer is similar to features seen
in simulations of shell galaxies (e.g., Quinn \cite{quinn84}, 
Hernquist \& Quinn \cite{hernquist88}) and 
could be interpreted 
as due to the infall of a dynamically colder galaxy $\sim$0.5 Gyr ago. 
Mackie \& Fabbiano (\cite{mackie98}) have argued from the temperature of the hot gas detected in the 
L$_1$ region that L$_1$ could be the
remnant of a small gas-rich galaxy on a high-velocity encounter 
($\sim$380 km s$^{-1}$) with NGC~1316, and its origin would therefore be different from that
of the other loops which formed in the aftermath of the major, older merger. 
The extended emission-line region EELR is found in the area of highest surface 
brightness of L$_1$. 
The fact that we don't detect any CO from the EELR is not surprising if the infalling galaxy
was metal-poor. 
The low level of detectable \hi emission from that region may be due to the fact that the
most of the atomic hydrogen has been stripped, or photo-ionized by the newly formed OB associations.

The NW concentration could as well be a recently accreted gas-rich galaxy.
It has a linear dimension of about 2.5 kpc and contains 
about 2.2$\times$10$^8$ M$_\odot$ of molecular hydrogen.  
This H$_2$ mass is comparable to that of NGC~1317, 
the small ($\sim$15 kpc diameter) spiral companion of NGC~1316  
(Horellou et al. \cite{horellou95}). 
This gas concentration is located 40$''$ (3.5 kpc) from the nucleus, 
slightly to the north of the minor axis. The velocity of the molecular gas, 
$v \sim$1850 km s$^{-1}$, is $\sim$70 km s$^{-1}$ higher than the systemic
velocity given by Arnaboldi et al. (\cite{arnaboldi98}) and agrees exactly 
with the stellar velocity derived from absorption line spectra in that region 
(figure~3 of Arnaboldi et al. \cite{arnaboldi98}).  Interestingly, beyond 
30$''$ from the nucleus along the minor axis, the stellar radial velocities increase 
on both sides, producing a U-shaped velocity curve, and the velocity dispersion 
profile is irregular. Arnaboldi et al.  argue that this kinematical feature is
caused by infalling material within one dynamical time-scale at 40$''$ (i.e. 
less than 3 $\times$ 10$^7$ years ago), since such features are not expected 
to last longer. 

\subsection{Gaseous shells?}

It is interesting to compare NGC~1316 with NGC 5128 (Centaurus~A). 
NGC 5128 lacks tidal tails but it 
has a more prominent dust lane and a better defined system 
of shells in both of which atomic and molecular gas have been detected
(Eckart et al. \cite{eckart90}; 
Schiminovich et al. \cite{schiminovich94}; 
Charmandaris et al. 
\cite{charmandaris00}).  
NGC 5128 has comparable amounts of \hi and H$_2$ in the region of the dust lane 
($\sim 3.5\times 10^8$ M$_\odot$) whereas NGC~1316 has at least four times more 
molecular gas than \hi.  
This difference in \hi content may be due to the fact that the \hi in NGC~1316 
is more affected
by the X-ray emission, since 
NGC~1316 is brighter in X-ray than NGC 5128 
(factor of 8 in the 0.1--2.4 keV band, ROSAT observations, and
factor of 1.4 in the 0.2--4 keV band, $Einstein$ $IPC$ observations).  
Both galaxies have 
a double-lobe radio source. The radio distribution, however, has a 
different orientation: in Centaurus~A it is perpendicular to the dust lane 
whereas in Fornax~A it is not. Nuclear effects may therefore affect the gas and dust properties
more clearly in Fornax~A than in Centaurus~A. 
In the outer parts of NGC 5128, \hi is found in the shells, with CO detected in 
the highest column density regions with M(\hi)/M(H$_2$)$\sim$1. 
In NGC~1316, we could not detect any CO from the bright outer \hii regions, but
our limit on the H$_2$ mass is comparable to the \hi mass. 
The possible chain of \hi clumps in the outer part of NGC~1316 
could be the highest peaks in a continuous shell. More sensitive observations are
required to confirm this. 

The $HST$ image of NGC~1316 displays beautifully the distribution of the dust
(Fig.~11c).
Northwest of the nucleus it roughly has a T-shape with a radial extent of 
about 40$''$ and an azimuthal extent of almost 90$^{\circ}$. 
CO is detected in the area of both the radial and the azimuthal dust patches. 
Grillmair et al. (\cite{grillmair99})  point out the similarity of the radial 
structure to the photo-evaporating columns in the Eagle Nebula M16 
(Hester et al. \cite{hester96}), though on a much larger scale, and suggest 
that it may be due to Rayleigh-Taylor instabilities or subject to erosion
by particles or radiation from the nucleus of the galaxy.  
The azimuthal feature may be a dusty shell.
The azimuthal configuration is expected from recently accreted material 
oscillating between two apocenters and forming shells (e.g. Hernquist \& Quinn 
\cite{hernquist88}). Gaseous shells could form in the same phase-wrapping process 
if the gas is clumpy and not very dissipative 
(Charmandaris \& Combes \cite{charmandaris-turku}). 
A detailed dynamical model of the gaseous features in NGC~1316 based on the accretion
of a small companion will be the subject of another paper.  

\section{Summary and conclusions}

Our new observations of molecular gas (as traced by $^{12}$CO(1--0) and $^{12}$CO(2--1) line
emission) toward the peculiar elliptical galaxy NGC~1316 
and a new analysis of \hi data have shown that: 

\begin{enumerate}

\item{About 5$\times$10$^8$ M$_\odot$ of molecular gas is located in the inner 2$'$ 
of the galaxy and is mainly associated with the dust patches along the minor axis 
($\sim$1.8$\times$10$^8$ M$_\odot$ in the southeast, $\sim$10$^8$ M$_\odot$ 
in the center and $\sim$2.2$\times$10$^8$ M$_\odot$ in the northwest).  
Those mass estimates are based on a Galactic CO/H$_2$ conversion factor. }

\item{No neutral atomic gas found was found in the spheroid of NGC~1316, down to a limit 
of $\sim$10$^8$ M$_\odot$. 
\hi was detected at four locations in the outer part of NGC~1316. }

\item{Two of those locations coincide with bright \hii regions in a tidal loop. 
About 2$\times$10$^7$ M$_\odot$ of \hi was found toward the giant \hii region identified by Schweizer 
(\cite{schweizer80}) and located 6$\farcm$7 (or 36.2 kpc) from the nucleus of NGC~1316. 
No CO could be detected. 
About 10$^7$ M$_\odot$ of \hi was detected toward the extended emission line region studied  
by Mackie \& Fabbiano (\cite{mackie98}). }

\item{The NW molecular gas concentration could be due to recent infall of material
onto NGC~1316. It could be a small gas-rich galaxy or the trace of a gaseous
shell formed though radial oscillations of infalling material.}

\item{The two spiral companions of NGC~1316, NGC~1317 to the north and NGC~1310 to the 
east, are located within the primary beam of the VLA and their \hi distribution and 
kinematics could be mapped. It was found that NGC~1317 has an unusually small
\hi disk that may have been affected by ram-pressure stripping if it passed through
the X-ray emitting region of NGC~1316. The \hi distribution of NGC~1310 is asymmetric.}

\item{There is a hint that the central radio jet of Fornax~A is bent owing to 
interaction with the large concentrations of interstellar gas seen in CO.}

\end{enumerate}

\begin{acknowledgements}
We are grateful to the SEST staff for their help and support during
the observations and to the VLA analysts for their prompt replies and useful
help.  
This work was in part supported by an NSF grant to Columbia University.
We have used the tool DEXTER provided by NASA Astrophysics Data System
to extract data from Fig.~11a of Schweizer (\cite{schweizer80}) and 
from Fig.~2 of Geldzahler \& Fomalont (\cite{geldzahler84}). 
Those figures and the radio continuum map of Fomalont et al. (\cite{fomalont89}) 
shown in Figure~1 were used with the permission of the authors. 
We have used images from the Digitized Sky Surveys. 
The Digitized Sky Surveys were produced at the Space Telescope Science
Institute under U.S. Government grant NAG W-2166. The images of these
surveys are based on photographic data obtained using the Oschin Schmidt
Telescope on Palomar Mountain and the UK Schmidt Telescope. 
We are grateful to the referee for a careful reading of the manuscript 
and useful comments. 
\end{acknowledgements}

\end{document}